%% file: main.tex
\documentclass[conference,10pt,top=0.7in]{IEEEtran}
\usepackage{graphicx,color}%,spconf}
\usepackage{amsmath, amsthm, amsfonts, amssymb, amsbsy,nccmath}
\usepackage{mathtools}

\usepackage{algorithm}% http://ctan.org/pkg/algorithms
\usepackage{algcompatible}% http://ctan.org/pkg/algorithmicx
\usepackage{enumerate}
\usepackage{lipsum}
\newtheorem{lemma}{Lemma}
\usepackage{textgreek}
\usepackage{textcomp}

\usepackage{algpseudocode}% http://ctan.org/pkg/algorithmicx

\usepackage[sort,compress]{cite}
\usepackage{epsfig}
\usepackage{epstopdf}
\usepackage{mathtools}
\usepackage{dsfont}
\usepackage{epstopdf}

\usepackage{sidecap, caption}
\usepackage{subcaption}
\theoremstyle{definition}
\theoremstyle{assumption}
\newtheorem{assumption}{Assumption}[section]

\theoremstyle{proposition}

\theoremstyle{corollary}

\usepackage[inline]{enumitem}   
\makeatletter
\newcommand{\inlineitem}[1][]{%
\ifnum\enit@type=\tw@
    {\descriptionlabel{#1}}
  \hspace{\labelsep}%
\else
  \ifnum\enit@type=\z@
       \refstepcounter{\@listctr}\fi
    \quad\@itemlabel\hspace{\labelsep}%
\fi}
\makeatother
\newcommand\norm[1]{\left\lVert#1\right\rVert}

\input{defines}
\makeatletter
\newcommand\fs@spaceruled{\def\@fs@cfont{\bfseries}\let\@fs@capt\floatc@ruled
  \def\@fs@pre{\vspace{0.5\baselineskip}\hrule height.8pt depth0pt \kern2pt}%
  \def\@fs@post{\kern1pt\hrule\relax}%
  \def\@fs@mid{\kern2pt\hrule\kern2pt}%
  \let\@fs@iftopcapt\iftrue}
\makeatother
\newcommand{\bit}{\begin{itemize}}
\newcommand{\eit}{\end{itemize}}

\newcommand{\mK}{\mathcal{K}}
\newcommand{\mL}{\mathcal{L}}

\newcommand{\mG}{\mathcal{G}}
\newcommand{\mC}{\mathcal{C}}

\newcommand{\bmu}{\boldsymbol{u}}
\newcommand{\bmz}{\boldsymbol{z}}

\newcommand{\bmg}{{\boldsymbol{g}}}

\newcommand\blfootnote[1]{%
  \begingroup
  \renewcommand\thefootnote{}\footnote{#1}%
  \addtocounter{footnote}{-1}%
  \endgroup
}

\newcommand{\bphi}{\boldsymbol{\phi}}

\newcommand{\bms}{{\boldsymbol s}}

\DeclarePairedDelimiter\abs{\lvert}{\rvert}%

\newcommand{\bmsh}{\widehat{\bms}}

\usepackage{lipsum}
\makeatletter

\newcommand\longleftrightarrowfill@{%
  \arrowfill@\leftarrow\relbar\rightarrow}
\makeatother

\restylefloat{algorithm}
\usepackage{flushend}
\begin{document}
%\linespread{0.99}
\title{ Reasoning with the Theory of Mind for Pragmatic Semantic Communication\vspace{-4.5mm}}
\vspace{-8mm}
\author{\fontsize{1}{1}\selectfont
\IEEEauthorblockN{\fontsize{11}{12}\selectfont
Christo Kurisummoottil Thomas\IEEEauthorrefmark{1}, and Emilio Calvanese Strinati\IEEEauthorrefmark{2}, and Walid Saad\IEEEauthorrefmark{1} 
\vspace{0mm}}
%\IEEEauthorblockN{\fontsize{10}{12}\selectfont
%\IEEEauthorrefmark{1}\fontsize{9}{9}\selectfont Qualcomm, Finland and \\
\IEEEauthorrefmark{1}Wireless@VT, Bradley Department of Electrical and Computer Engineering, \\ Virginia Tech, Arlington, VA, USA, \IEEEauthorrefmark{2} CEA-Leti, Grenoble, France.\\ \fontsize{8}{8}Emails: \{christokt,walids\}@vt.edu,emilio.calvanese-strinati@cea.fr\vspace{-6mm}
}
\maketitle

\vspace{-0mm}
\begin{abstract}\vspace{-0mm}
In this paper, a pragmatic semantic communication framework 
 that enables effective goal-oriented information sharing between two-intelligent agents is proposed. In particular, semantics is defined as the causal state that encapsulates the fundamental causal relationships and dependencies among different features extracted from data.
The proposed framework leverages the emerging concept in machine learning (ML) called \textit{theory of mind (ToM)}. It employs a dynamic two-level (wireless and semantic) feedback mechanism to continuously fine-tune neural network components at the transmitter. Thanks to the ToM, the transmitter mimics the actual mental state of the receiver's reasoning neural network operating semantic interpretation. 
Then, the estimated mental state at the receiver is dynamically updated thanks to the proposed dynamic two-level feedback mechanism. At the lower level, conventional channel quality metrics are used to optimize the channel encoding process based on the wireless communication channel's quality, ensuring an efficient mapping of semantic representations to a finite constellation. Additionally, a semantic feedback level is introduced, providing information on the receiver's perceived semantic effectiveness with minimal overhead. 
Numerical evaluations demonstrate the framework's ability to achieve efficient communication with a reduced amount of bits while maintaining the same semantics, outperforming conventional systems that do not exploit the ToM-based reasoning.

\end{abstract}\vspace{-0mm}

\blfootnote{The work of Emilio Calvanese Strinati is supported by ``6G-GOALS", an EU-funded project, and the French project funded by the program "PEPR Networks of the Future" of France 2030. The work of Christo Kurisummoottil Thomas and Walid Saad was supported by the Office of Naval Research (ONR) under
MURI grant N00014-19-1-2621. }
\vspace{-3mm}
\section{Introduction}
\label{section_intro}
\vspace{-1mm}

Future 6G wireless systems will transmit and process large volume of data potentially under stringent requirements of ultra-low latency, ultra-reliability and high throughput, to support new \emph{connected intelligence (CI)} applications such as haptics, brain-computer interaction, flying vehicles, extended reality (XR), and the metaverse \cite{ChaccourITJ2022}. 6G systems could transmit only essential information relevant to the end-user to meet this need for high-rate, high-reliability, and time-criticality for the aforementioned CI applications, . This concept forms the basis of \emph{semantic communication (SC)} systems \cite{ChaccourArxiv2022,StrinatiComNetworks2021,UysalIN2022}.

{Despite a recent surge in SC research \cite{XieTSP2021, FarshbafanArxiv2022}, focus has been mainly on potential SC gains assuming an error-free SC channel. Nevertheless, no channel is error-free. With SC, the communication channel has two components: wireless and semantic. One critical challenge that often goes unnoticed is the need to dynamically adapt transmission strategies at the semantic level based on end-to-end performance.
The evaluation of errors at the semantic channel was provided by \cite{Sana2022} for the first time, showing that the semantic channel's quality critically impacts the overall communication performance. 
In wireless communication, adaptive mechanisms adjust transceiver design based on estimated link quality for a trade-off between robustness and efficiency. This paper asserts the need for adaptive mechanisms at the semantic level as well \cite{CalvanesePatent2020}.
This requires at the system architecture level the introduction of a \textit{semantic plane} \cite{Petar2020} where feedback information from the semantic receiver to the semantic transmitter can be exchanged over a dedicated plane.
%Despite a recent surge in SC research \cite{XieTSP2021,Sana2022,LiuISIT2021,Mehdi2021, FarshbafanArxiv2022,FarshbafanICC2022}, most of these prior works fall short of addressing key challenges needed to integrate into existing digital communication hardware. One critical challenge that often goes unnoticed is the need to adapt transmission strategies based on end-to-end performance dynamically. 
%{To minimize resource use and compensate deviation from effective semantic interpretation at the receiver, the overall SC system must be capable of dynamically adjusting to communication environment dynamics the semantic encoder and decoder components settings.  }
%To effectively harness the benefits of SC systems in practical wireless environments, wireless systems must be capable of dynamically adjusting the semantic encoder and decoder components. 
The adaptation of semantic encoder and decoder should be driven by the receiver's performance in reconstructing semantic information, ensuring optimized communication and improved overall system efficiency. 
%% I would keep this sentences only if we have room. there is much selling and not muchnew concepts here :) ####  BLUE color to indicate that we can skip the sentence. ####
{By addressing these challenges, we can pave the way for the successful integration of SC systems into existing digital communication infrastructure, with minimal changes to existing layered architecture.} 
{However, many existing works \cite{XieTSP2021, Sana2022, FarshbafanArxiv2022} 1) fix  the parameters of transmitter and receiver neural network (NN) components during offline training  and, 2) NN's parameters are jointly learned to minimize semantic reconstruction errors. This approach is often adopted due to the increased complexity and re-training overhead to update the parameters via online learning.} 
%
%Existing works in SC, such as \cite{XieTSP2021, Sana2022, LiuISIT2021, Mehdi2021, FarshbafanArxiv2022, FarshbafanICC2022}, often rely on fixed parameter settings for transmitter and receiver NN components determined through offline training. This approach is often adopted due to the increased complexity and re-training overhead to update the parameters via online learning. 
The training aims to {minimize semantic reconstruction errors while maximizing a performance target such as semantic compression gain.} {Nevertheless, such approach to training} does not guarantee {expected semantic reliability \cite{ChristoTWCArxiv2022}} in dynamic communication scenarios where environment variations or {task's specifications} changes may occur. %Furthermore, many current SC system designs, particularly those based on an end-to-end framework, commonly utilize analog modulation, exemplified by \cite{XieTSP2021} and \cite{ShaoJSAC2022}. However, a few exceptions, such as \cite{WangSPL2022} and \cite{TungJSAIT2022}, consider finite constellation designs for SC systems. Even in these designs, the dynamic adaptation of deep learning components to accommodate changes in channel environments or user intents needs to be explicitly addressed. 
Herein, a key concern arises: \textit{how should the semantic design components be adjusted to avoid or rectify {errors at the semantic level} %semantic errors 
in subsequent transmissions if the receiver cannot decode the information reliably?} This question draws parallels to the concept of automatic repeat request (ARQ) 
%or hybrid ARQ 
used in traditional systems {where a feedback control channel is used to inform on the positive or negative acknowledgement on received information. In an SC system, transmitters and receivers must extract (at the transmitter) and interpret (at the receiver) semantic messages based on different beliefs and knowledge. Such mismatch on logic and reasoning states at transmitter and receiver can cause critical errors, which most of the time are undetected and thus systematically repeated}. %

We propose harnessing the idea of ``\textit{theory of mind}" (ToM) \cite{YuanArxiv2021} to {tackle these challenges by operating before communication at the transmitter an estimate of the mental state of the receiver (a priori knowledge and believes). Thanks to such a pragmatic communication approach, we contribute to improving the overall SC communication accuracy, reliability, and reduction of needed communication and processing resources.}

\subsection{Contributions}

%In contrast to existing approaches, 
%{The main novelty of this paper is the definition of a comprehensive framework that enables ToM reasoning in point-to-point SC systems.} 
{Our framework considers the semantic information present in the data to be the causal state extracted using graph NNs (GNN) \cite{ScarselliGNN2008} from observed time series data at the transmitter}. We chose GNN since it provides a rigorous framework to extract the causal relations present among the features in the data as a directed acyclic graph. 
%{In our work we assume semantic information extracted from  observed time series data at the transmitter to have intrinsic causal relations. Specifically, we adopt GNNs \cite{ScarselliGNN2008} to extract their intrinsic causal states.}
By leveraging ToM, we can dynamically adapt the parameters of NNs based on feedback measures of semantic effectiveness. This approach enables effective semantic design component adjustment, improving subsequent transmissions' reliability and performance in SC systems.  Additionally, to accurately reason about other nodes, we introduce a two-level feedback mechanism. The first level involves a semantic level of feedback, enabling dynamic adjustments to the semantic representation component and causal state extraction. The second level consists of conventional channel quality feedback, used to optimize the design of the constellation-constrained channel input. %Overall, our framework combines ToM reasoning and two-level feedback to enhance the semantic reliability and adaptability of point-to-point SC systems, with reduced overhead compared to state-of-the-art. 
Finally, we demonstrate via simulations the superior spectral efficiency of the proposed ToM-based SC system compared to conventional schemes such as repetition coding and HARQ applied to SC systems. Additionally, we demonstrated that the proposed SC system requires fewer redundant semantic symbols to be transmitted in order to achieve close to perfect semantic reliability. This is in contrast to the state-of-the-art systems that do not employ ToM reasoning. 

\vspace{-0mm}\section{System Model}

\begin{figure*}[t]
%\vspace{-1mm}
%\centerline{\includegraphics[width=8.3cm,height=5cm]{SSP_NMSE_GAMP_IF_SAVE}}
%\centerline{\includegraphics[width=9cm]{NMSE_EM_LAMP_SBL}}
\centerline{\includegraphics[width=6.8in,height=2in]{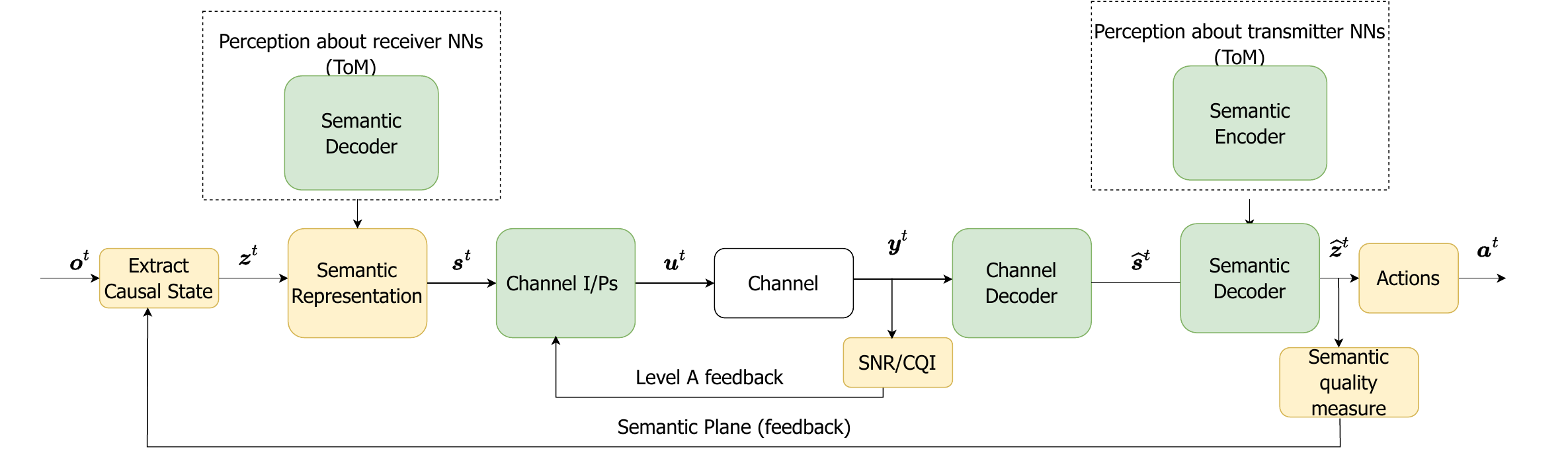}}\vspace{-3.5mm}
\caption{\small Proposed system model.}
\label{SCModel}\vspace{-0mm}
\vspace{-3mm}
\end{figure*}\vspace{-1mm}
\begin{figure*}[t]
\beq
\vspace{-0mm}
\begin{array}{l} 
C_t(\bms^t,\bmsh^t)= \textrm{KLD} \left(\pi(\bma^t \mid \bmz^t) \,\mid\mid\, \sum\limits_{\bmy^t,\bmu^t,\bmsh^t} p({\bmy}^t \mid {\bmu}^t )p(\bmsh^t\mid\bmy^t)p(\widehat{\bmz}^t \mid {\bms}^t )\pi(\bma^t\mid\widehat{\bmz}^t)\right).
\end{array}
\label{eq_sem_eff}
\vspace{-3mm}
\eeq
\end{figure*}
We consider the SC system shown in Fig.~\ref{SCModel}. In this system, a transmitter has access to a sequence of observations represented as a time series data, where each $N-$dimensional vector of observations $\bmo^t$ belongs to a set $\mathcal{O}$. The transmitter's goal is to transmit a concise description, $\bmz^t$, of the data $\bmo^t$ that can be efficiently interpreted at the receiver. This description captures the semantic information of the data, including the intrinsic causal relationships \cite{ChaccourArxiv2022}. The receiver aims to use the reconstructed received messages $\bmzh^t$ to perform a set of actions $\bma^t \in \mA$, sampled using the distribution $\pi(\bma^t\mid\bmzh^t)$. Contrary to previous works \cite{XieTSP2021}, we constrain the channel input to a finite constellation. As such, the channel inputs $\bmu^t$ are selected from a digital modulation scheme of order $M$, where the constellation symbols are represented by $\mM = \{c_1,\cdots,c_M\}$. Consequently, the channel input $\bmu^t$ is a digital sequence of varying dimension (that gets learned), with each element drawn from the constellation symbols $\mM$. Hence, we refer to this sequence as a \emph{constellation symbol sequence}. 

The primary innovation that distinguishes this paper from the state-of-the-art \cite{XieTSP2021, Sana2022, FarshbafanArxiv2022} is the incorporation of receiver feedback, denoted as $d^t \in \mathcal{D}$ (a scalar value), through a feedback channel called the ``semantic plane." The set $\mD$ here represents the \emph{semantic effectiveness} measure. Semantic effectiveness captures the accuracy of the actions performed at the receiver using the reconstructed semantic state $\bmzh^t$. In particular, to capture the semantic effectiveness, we introduce the metric $C_t$, which captures the causal impact of the transmitted message (via the receiver's actions) as observed through a channel with a response characterized using $p(\bmy^t\mid \bmu^t)$ (this distribution could capture the fading and interference in the wireless environment). In other words, $C_t$ measures the semantic effectiveness (inversely proportional to $C_t$) of the transmitted message to the end-user and is written as \eqref{eq_sem_eff}, where $\textrm{KLD} (p\mid\mid q)$ represents the KL divergence between $p$ and $q$. Hence, the feedback $d^t$ gives the transmitter an indication of whether those are accurate in terms of achieving the system's long term goal. The transmitter can adjust its transmission strategy and update its causal reasoning components by using the feedback received through the semantic plane. This dynamic nature of our proposed method ensures that the NN components are continuously updated based on the real-time feedback from the receiver. As a result, the system can adapt and optimize its performance dynamically during communication, leading to more efficient and effective semantic communication. We also consider a second level of feedback through classical measures like the signal-to-noise-ratio (SNR) or the channel quality indicator (CQI) measure, defined as $p^t$, which gets used by the channel encoder to dynamically adjust its constellation sequence.  

As an illustrative example of the proposed system, we consider a variant of the problem of rendezvous between two robots exploring an unknown environment \cite{DudekAAAI2009}.
That is, how can two autonomous exploring agents
that can communicate with one another over
long distances meet if they start exploring at different locations in an unknown environment. The intended application is collaborative map exploration. Herein, we can consider the case in which only one of the robot, say $A$ has the necessary sensing modalities to detect the landmarks near it. Robot $A$ communicates this information to robot $B$, which receives it over a wireless channel. Moreover, utilizing the extracted information, $B$ attempts to make additional movements along the trail and feedback its actions back to robot $A$. Next, we define the causal generative model that describes the data generation and transmission signal.
\vspace{-2mm}\subsection{Causal discovery model}
\label{CausalDiscovery}\vspace{-1mm}

The causal relationships present within the source data can be represented as a causal graph, and the process of learning this graph is referred to as the \emph{causal discovery} problem \cite{BareinboimACM2020}. In a complex environment with high-dimensional observations, such
as images or video, learning a compact latent state space that captures the causal dynamics of the environment has
been shown to be more computationally efficient than learning predictions directly in the observation
space \cite{HafnerPMLR2019}. This latent space, denoted $\bms^t$ is the \emph{semantic representation} here.
The generative model for a dataset of observations and  transmitter actions ($\bmu^t$), represented as $(\bmo^{0:T}, \bmu^{0:T})$, can be formulated as follows ($\btheta$ is the set of the NN parameters):
\vspace{-0mm}\beq
\vspace{-0mm}
\begin{aligned}
p(\bmo^{0:T}, \bmu^{0:T}) = &\int \prod\limits_t   p_{\btheta}(\bmo^t\mid \bmz^t) p_{\btheta}(\bmz^t\mid\bmz^{t-1},d^{t-1}) \\ &\pi_{\btheta}(\bms^t\mid\bmz^{t})d\pi_{\btheta}(\bmu^t\mid\bmz^{t},p^{t-1})d\bmz^{0:T}d\bms^{0:T},
\end{aligned}\label{eq_genModel}
\vspace{-0mm}\eeq
where $p_{\btheta}(\bmo^t\mid \bmz^t)$ and $p_{\btheta}(\bmz^t\mid\bmz^{t-1},d^{t-1})$ are the observation model and the causal transition model, respectively. %We explicitly impose the constraint that the scalar variables in $\bmo^t$ are causally related, and this causal structure ($\bmz^t$ of dimension $N_z$) needs to be learned. 
The causal structure $\bmz^t$ is represented as a sequence of entities (scalar variables present in $\bmo^t$) and their relations. While extracting the causal structure, any variables in $\bmo^t$ that remain unconnected in the graph can be considered irrelevant and will not be transmitted.
As mentioned in previous studies \cite{ChristoTWCArxiv2022,ChristoJSAITArxiv2023}, acquiring causal representations can result in AI solutions that exhibit \emph{distribution invariance}, enabling adaptability across various interventional distributions. 
In this paper, we opted not to delve into discussions about interventions as they have already been addressed in our previous work \cite{ChristoTWCArxiv2022}. From SC perspective, this implies constructing solutions with reduced training overhead, leading to improved real-time decision-making capabilities compared to data-intensive AI approaches. It is important to note that none of the factors in equation \eqref{eq_genModel} are known and must be learned from the data. The feedback of the receiver actions $d^t$ is reflected in the transmitter policy $\pi_{\btheta}(\bms^t\mid\bmz^{t})$ that chooses the transmitted semantic representation, at time $t$. The design of the representation $\bms^t$ relies on the specific semantic information intended to be conveyed, which, in turn, is influenced by the receiver's actions. Hence, we next define the semantic information $\mbS_t$, essential for the subsequent problem formulations.

\vspace{-1mm}\subsection{Semantic information measure}

Using the novel semantic information measures in \cite{ChristoTWCArxiv2022}, we directly obtain the transmitted and received semantic information as follows. We omit the time superscript for convenience.
\vspace{-0mm}\begin{lemma}
\vspace{-0mm}\label{theorem_semInf}
The \emph{semantic information} conveyed by any causal state $\bmz\in \mC_1$ can be written as the average across the information carried by the possible actions ($\bma$) that follow from $\bmz$.  Further, mathematically, we can write the semantic information conveyed by $\bmz$ as
\vspace{-1mm}
\beq
\vspace{-2mm}
                           \mathbb{S}(\bmz) = \sum\limits_{\bma \in \mA} \pi(\bma\mid \bmz)  \log \frac{\pi(\bma\mid \bmz) }{\pi(\bma)} = \textrm{KL}(\pi(\bma\mid \bmz)\mid\mid \pi(\bma)).
\eeq
\vspace{-2mm}
\label{eq_SemInfMeasure_x_c_d}
\end{lemma}
Leveraging \cite[Corollary 1]{ChristoTWCArxiv2022}, we obtain the average semantic information extracted by the receiver as \eqref{eq_list_semInfo}. $Z_{\bmzh_t\bmz_t}$ is the semantic similarity between the transmitted and reconstructed causal states, which is defined in \cite[Corollary 1]{ChristoTWCArxiv2022}.
\begin{figure*}
\begin{align}\mathbb{E}_{\bmu_t}\left[S_{r,t}(\bmzh_{k,t};\bmu_{k,t}\mid l_{k,t},[\bmzh_{k,t-1}],\pi_{l,t}) \right]   & = \\  &\hspace{-10mm} \sum\limits_{\bmu_{t},\bmy_{k,t}}\!\!\!\!\pi\left(\bmu_{k,t}\mid [\bmz_t]\right)S({\bmz}_{t}\mid [\bmz_{t-1}])p(\bmy_{t}\mid \bmu_{t})\left[\sum\limits_{\bmzh_{t}}\pi_{\bmz_t}\log\frac{\pi_{\bmz_t}}{\pi(\bmzh_{t}\mid\bmzh_{t-1}])}Z_{\bmzh_t\bmz_t}\right]. \label{eq_list_semInfo}
\end{align}\vspace{-3mm}
\end{figure*}

\vspace{-2mm}\subsection{Semantic reliability}\vspace{-2mm}

\emph{Semantic reliability} is quantified by the expression, 
$
p\left(E_t\left({\bmz}^t,{\widehat{\bmz}}^t \right) < \delta \right) \geq 1-\epsilon,
$
where $E_t(\bmz^t,{\widehat{\bmz}}^t) = \norm{\bmz^t-\bmzh^t}^2$ represents semantic distortion and $\epsilon$ is arbitrarily small. This definition follows our previous works \cite{ChristoTWCArxiv2022,ChristoJSAITArxiv2023}. This metric reflects the receiver's ability to reliably reconstruct all the causal aspects in the decoded causal structure. Unlike classical reliability measures, in semantics, we can recover the actual meaning of transmitted messages even with a higher bit error rate (BER), as long as the semantic distortion remains within the set limit. This is illustrated by the choice of $\delta$ here, which depends on the concept of semantic space. The \emph{semantic space} $\mK$ is defined as an $N$-dimensional topological ball centered at the actual state $\bmz^t$, where all points inside the ball corresponding to the same semantic information. Formally, we can express this as $E_t(\bmz^t,{\widehat{\bmz}}^t) \leq \delta$ such that $\mbI_{\phi}(\bmz^t)=\mbI_{\phi}(\bmzh^t)$, where $E_t$ is the topological distance between states $\bmz^t$ and $\bmzh^t$, $\delta$ is the radius of the topological ball, and $\mbI_{\phi}$ is a mapping function that maps states to their corresponding semantic information. 

Having defined the necessary metrics, we next formulate the problem for the transmitter and receiver components in our SC system.

\vspace{-2mm}\section{Theory of Mind based SC Design}

Here, we first discuss the causal discovery component at the transmit side, that represents the mapping from $\bmo^t$ to $\bmz^t$.

\vspace{-1mm}\subsection{Causal discovery at the transmitter}\vspace{-1mm}

The fundamental assumption underlying causal discovery here is that there exists a fixed function $\bmg$ that characterizes the dynamics of all samples $\bmo^t \in \mO$, given their past observations up to time $t$ and the underlying causal graph $\mG$. This can be written using the Granger causality concept \cite{GrangerJES1969} as follows:
\beq
\bmo^{t+1} = \bmg(\bmo^{\leq t},\mG) + \bmn^{t+1}.
\eeq
In this data-generating process, we model two variables: the causal graph $\mG$, and the dynamics $\bmg$, which are shared across all samples. By separating the causal graph from the dynamics, we can design our model accordingly. Specifically, we introduce a causal discovery encoder $f_{\btheta}$, which learns to infer the causal graph $\mG$ given the sample $\bmo^t$. $\btheta$ represents the NN parameters. Additionally, we introduce a dynamics decoder $\bmg_\theta$ that learns to approximate the dynamics $\bmg$. This division of the model allows us to effectively capture the relationship between the causal graph and the shared dynamics within the data. 

\subsubsection{Problem formulation}

To approach the problem of causal discovery, we adopt a probabilistic framework and model the functions $f_{\btheta}$ and $g_\theta$ using variational inference. We choose the encoder $f_{\btheta}$ by incorporating an encoding function $q_{\btheta}(\bmz\mid\bmo)$, which produces a distribution over $\bmz$ that represents the predicted edges $\mE^\text{Enc}$ in the causal graph. Simultaneously, we train a decoder $p_{\theta}(\bmo \mid \bmz)$ that probabilistically captures the dynamics of the time-series based on the predicted causal relationships.

In our framework, we select the negative log-likelihood as the loss function $\mathcal{V}_f$ for the decoder. %, denoted as $\mathcal{L}_{\text{decoder}}$. 
Additionally, we incorporate a regularizer that measures the Kullback-Leibler divergence  (KLD) to a prior distribution over $G$. This regularizer encourages the model to learn causal structures consistent with our prior knowledge. Overall, our loss function $\mV_f$ represents a variational lower bound \cite{FoxAIR2012}:
\beq
\mV_f = \mathbb{E}_{q_{\phi}(\bmz\mid\bmo)} \log p_{\theta}(\bmo\mid\bmz) - \textrm{KL}(q_{\phi}(\bmz\mid \bmx)\mid\mid p(\bmz)).
\label{eq_vlb}
\eeq

%\noindent We can express \eqref{eq_vlb} as $-\mV_f \geq -\mathbb{E}_{q{\phi}(\bmz\mid\bmo)} \log p_{\theta}(\bmo\mid\bmz)$, which serves as an upper-bound on the \emph{transmitter surprise} when observing $\bmo$. The term $-\mV_f$ represents the variational free energy (VFE) \cite{FristonNRN2010}. In this context, the objective for the transmitter is to minimize the VFE, which can be viewed as minimizing the level of surprise. In simpler terms, the internal model representing the causality on the transmitting side should accurately align with the external observations to minimize surprise. 
Further, we formally write the causal discovery problem as (with variables $x^*$ representing the optimized quantities):
\beq
\left[\mG^*,\bmz_t^*\right] = \argmin\limits_{\mG} -\mV_f.
\eeq

\subsubsection{Causal discovery solution}

To optimize \eqref{eq_vlb}, The encoder $q_{\btheta}(\bmz\mid \bmo)$ is designed as a GNN \cite{ScarselliGNN2008}, denoted as $f_{\text{enc},{\btheta}}$. %This network operates on the input data and performs information propagation across a fully connected graph $\mG = \{\mV, \mE\}$. In this graph, each time-series $i$ is represented by a vertex $v_i \in V$, and there exists an edge connecting every pair of vertices $(v_i, v_j)$.
%\beq
%\begin{array}{l}
%\psi_{ij} = f_{enc,{\btheta}}(\bmo)_{ij}, \\
%q_{\btheta}(z_{ij}\mid \bmo) = \textrm{Softmax}%(\psi_{i,j}/\tau)
%\end{array}
%\eeq
The output $z_{ij}$ of the encoder represents the predicted edges $\mE^{enc}$ in the causal graph $\mG$.
%By leveraging the GNN, the encoder efficiently incorporates and propagates information across the connected vertices, allowing for the integration of relevant dependencies between different time-series. This facilitates the learning and inference of the causal graph structure within the causal discovery framework. 
The decoder $p_\theta(\bmo \mid \bmz)$ captures the dynamics of the time-series data based on the predicted causal relations. It takes into account both the predicted causal relations $z_{ij}$ and $\bmo^t$ as its input. First, it propagates information along the
predicted edges by applying a NN $f_e$, and compute $
h_{ij}^t =  \sum\limits_{e > 0} z_{ij,e} f_e([o_i^t,o_j^t])
$.
Then, the decoder accumulates the incoming messages to each node and applies a NN $f_v$
to predict the change between the current and the next time-step:
\beq\begin{array}{l}
\mu_j^{t+1} = o_j^t + f_v(\sum_{i\neq j}h_{ij}^to_j^t)\\
p_{\theta}(o_j^{t+1}\mid\bmo^t,\bmz) = \mN(\mu_j^{t+1},\sigma^2).
\end{array}
\eeq
\begin{figure}[t]
%\vspace{-1mm}
%\centerline{\includegraphics[width=8.3cm,height=5cm]{SSP_NMSE_GAMP_IF_SAVE}}
%\centerline{\includegraphics[width=9cm]{NMSE_EM_LAMP_SBL}}
\centerline{\includegraphics[width=3.5in,height=1.4in]{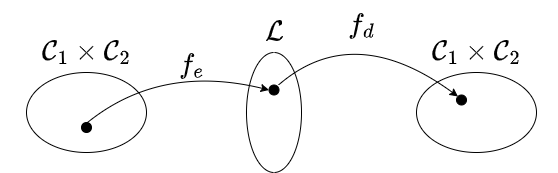}}\vspace{-1.5mm}
\caption{\small A semantic decoder function maps points in $\mC_1 \times \mC_2$ to $\mC_1 \times \mC_2$ via $\mL$ (the semantic language), where $f_e$ is the encoder
function and $f_d$ the decoder function.}
\label{ConceptualSpaces}\vspace{-0mm}
\vspace{-5mm}
\end{figure}\vspace{-0mm}
Having discussed the causal discovery problem, we next move on to computing the semantic representation $\bms^t$.

\subsection{Theory of mind based semantic representation}

 Communication is successful when the concept defined as $\bmz^t \in \mC_1$  is perceived correctly by the receiver. This means that the extracted causal state at the receiver, $\bmzh^t\in \mC_2$ is such that it belongs to the semantic space (defined as the Euclidean space over which the semantic information conveyed by $\bmzh$ is the same within a ball of radius $\delta$) \cite{ChristoTWCArxiv2022}. $\mC_i$ represents the conceptual space that includes the set of all possible meanings or structure present in the data \cite{WarglienSynthese2013}. The decoded state description at the listener, constrained to a \emph{semantic message space} (represented as $\mathcal{S}_{z^t}$ corresponding to message $\bmz^t$), can convey the same semantic content (defined $S_{z^t}$). As a result, they are considered \emph{semantically similar}. Hence, we can achieve the same \emph{semantic reliability}. This semantic message space corresponding to $\widehat{\bmz}^t$ can be described as
\beq
\vspace{-1mm}E(\bmz^t,{\widehat{\bmz}}^t)  \leq \delta,\,\, \mbox{s.t.}\,\, E({\mbS}_t,{\widehat{\mbS}}_t) = 0,
\vspace{-0mm}\eeq
where $E(\bmx,\bmxh) = \norm{\bmx-\bmxh}^2$. $\mbS_t$ denotes the semantic information conveyed by $\bmz^t$. From an SC point of view (see Fig.~\ref{ConceptualSpaces}), the aim is to ensure that the pair of state descriptions $(\bmz^t,\bmzh^t) \in \mC_1\times \mC_2$ satisfy the following conditions that represent a meeting of minds between transmitter and receiver,
\beq
\mL(\bmz^t,\bmzh^t) = (\bmz^t,\bmzh^t).
\eeq
The language represented by $\mL$ is a cascade of two functions, $\mL = f_d \circ f_e$, where 1) the encoder function $f_e$ defined as $\bms^{t\,*} = \argmax\limits_{\bms^t} \pi_{\btheta}(\bms^t\mid\bmz^{t},d^{t-1})$, and 2) the decoder function $f_d$ defined as $\bmzh^{t\,*} = \argmax\limits_{\bmz^t} p_{\bphi}(\bmz^t\mid\bmsh^{t})$. Further, we make the following assumption on the transmitted symbol.

\begin{assumption}
\label{assumption_1}
 For every state description $\bmz^t$, extracted at time $t$, only one semantic symbol vector $\bms^t$ is computed. Further, the feedback ($d^t$) about the actions executed at the receiver decide the future semantic states to be transmitted (indirectly represent retransmissions at the semantic level). 
\end{assumption}

\noindent The semantic level of feedback $d^t$ %introduces a novel concept, representing a measure of semantic effectiveness in the SC system. It 
serves as an indicator of the system's overall performance, encompassing end-to-end aspects. By incorporating the semantic level of feedback, the requirement for multiple layers of feedback found in conventional digital communication systems can be mitigated, reducing control information overhead.

%Let $\mC$ denote a small category, called a \emph{schema}. A $\mC$-set, also known as a copresheaf on $\mC$, is a functor
%from $\mC$ to the category \emph{Set}. The schema mentioned here refers to a category that serves as a framework for organizing types (data domains) and their functional relationships. In this category, objects represent types, and morphisms represent various relationships, such as "is-a" relationships and other functional connections, between these types. The $\mC$-Set specifically refers to the category of sets and functions. Consequently, a $\mC$-Set is a functor that maps types to sets and type relationships to functions. This means that for each type in the category, there is a corresponding set, and for each type relationship, there exists a corresponding function that operates on the corresponding sets. The causal state $\bmz^t$ can be represented as an object in $\mC$-Set. Actions are specified as spans in $\mC$-Set.
\subsubsection{Problem formulation at the transmitter}

Given the causal state $\bmz^t$, we further discuss how to compute the mapping from $\bmz^t$ to the semantic symbol $\bms^t$. Here, we assume that transmitter states can be decomposed into physical states and private receiver states, namely, $\widetilde{\mZ}_s = \mC_1\times \Omega_r$, where $\mC_1$ is the transmitter state space ($\bmz^t$ belongs here), $\Omega_r$ is the receiver's NN parameter space. Each element part of $\Omega_r$ is denoted as $\bomega_{r,i}$. For
transmitter, it performs according to its value function, $Q_s:\widetilde{\mZ}_s \times \mA_s \rightarrow \mR$, where $\mA_s$ represents the action space (the semantic symbol set for transmitter). We can define the value of a state as $V^{\pi}(\bms^t) = \mbE_{\bms^t,d^{t+1},\cdots}\sum\limits_t \gamma^t r(d^t,\bms^t)$, where $d^t = \frac{1}{1+C_t(\bms^t,\bmsh^t)}$, and $r(d^t,\bms^t)$ is the reward obtained upon performing the control actions $\bma^t$ by the receiver. Here, the semantic effectiveness measure fed back from the receiver represents the state transitions akin to reinforcement learning (RL). Further, we can also define the Q-function as $Q_s(d^{t},\bms^t) = r(d^t,\bms^t) + \gamma\mbE_{d^{t+1}\sim p(d^{t+1}\mid d^t,\bms^t)}\left[V^{\pi}(\bms^{t+1})\right]$. We define $\tau_s$
as transmitter's observation and action history $\{d^0,\bms^0,\cdots,d^t,\bms^t\}$. The transmitter estimates a belief about the receiver's knowledge of the transmitted message (obtained through the respective NN parameters) and their private states. The transmitter must maintain a \emph{belief} about the private state of receiver, denoted as $b_s^t(\bomega_{r,i})$. The transmitter policy averaged over the receiver's belief distribution will be:
\beq
\vspace{-1mm}\begin{aligned}
\pi_s(\bms^t\mid\bmz^t,\bomega_s^t,\tau_s^t) &= \\ &\frac{e^{\left(\beta\sum\limits_{\bomega_{r,i}\in\Omega_{r}}b_s^t(\bomega_{r,i}\mid\tau_s^t)Q_s(\bms^t,\bomega_s,\bomega_{r,i})\right)}}{\sum\limits_{\bms^{t\,\prime}\in \mS_t}e^{\left(\beta\sum\limits_{\bomega_{r,i}\in\Omega_{r}}b_s^t(\bomega_{r,i}\mid\tau_s^t)Q_s(\bms^t,\bomega_s,\bomega_{r,i})\right)}},
\end{aligned}
\label{eq_sample_actions}
\eeq
where $\bomega_s$ represents the NN parameters of transmitter. The next challenge lies in how the transmitter can maintain its belief given its observation history. We can utilize counterfactual
reasoning in our belief update function \cite{YuanNeurIPS2019}. The transmitter traverses all possible private states $\bomega_{r}$  and estimates how likely the actions it observed are taken given a specific set of private agent states are the correct one. Then, it updates its belief using Bayesian rule:
 \beq
 \begin{aligned}
 b_s^t(\bomega_{r,i}\mid \tau_s^t) = P(\bomega_{r,i}\mid \tau_s^t,\bma^t) \propto \widehat{\pi}_{r}(\bma^t\mid \bms^t,\bomega_{r,i})b_s^{t-1}(\bomega_{r,i}),
 \end{aligned}\label{eq_b_prvtstate}
 \eeq
where $\widehat{\pi}_{r}$ is transmitter's estimation of receiver policy, learned in centralized training. As per Assumption~\ref{assumption_1}, the transmitter utilizes the fed back actions $\bma^t$ (denoted $d^t$) from the receiver to compute \eqref{eq_b_prvtstate}. To complete the ToM framework, the transmitter must maintain a belief about the receiver's private states, and conversely, the receiver must also estimate the transmitter's state. %When there are two appropriate actions to achieve the task, one of which can convey the transmitter's private state to others while the other reveals little information, the first action should be preferred. 
This is because maintaining accurate beliefs about each other's private states is crucial for effective coordination among agents. Utilizing the obverter technique \cite{YuanNeurIPS2019}, we let the transmitter holds a belief $\widehat{b}_{r}$ as the estimation of receiver's
belief about $\bomega_s$ (representing transmitter's NN parameters). All we need is a
belief update function, $f_{r}
: \Delta(\Omega_s) \times \mA_s \times \widetilde{\mZ}_s \rightarrow \Delta(\Omega_s),$ where $\Delta(\Omega_s)$ represents a distribution. Belief update function $f_r$ takes in the old belief, transmitter's action, the physical state and returns
a new belief as its new estimation of receiver’s belief over $\omega_s$. The ability to correctly infer receiver's private states from their
actions and predict others belief about oneself introduces ToM into our agents. Given the semantic representation $\bms^t$, which is a continuous signal, mapping to the nearest element in the constellation $\mM_t$ can be computed using a scheme discussed in \cite{TungJSAIT2022} and is beyond the scope of this paper. 

\subsubsection{Problem formulation for ToM based receiver design}

The generative model of the observations and actions sequence at the receiver side can be written as follows:
\vspace{-0mm}\beq
\vspace{-0mm}
\begin{array}{l}
p(\bmsh^{0:T}, \bma^{0:T}) = \\ \int \prod\limits_t   p_{\btheta}(\bmsh^t\mid \bmu^t) p_{\btheta}(\bmz^t\mid\bmz^{t-1},\bmsh^t) \pi_{\btheta}(\bma^t\mid\bmz^{t})d\bmz^{0:T},
\end{array}
\vspace{-0mm}\eeq
where we split the observation model, $p_{\btheta}(\bmsh^t\mid \bmu^t) = \int_{\bms^t,\bmz^t} p_{\btheta}(\bmy^t\mid \bmu^t)p_{\btheta}(\bmu^t\mid \bms^t)p_{\btheta}(\bms^t\mid \bmz^t)p_{\btheta}(\bms^t\mid \bmz^t)dsdz$. In other words, receiver also learn the transmitter semantic encoder $p_{\btheta}(\bms^t\mid \bmz^t)$,  the channel model $p_{\btheta}(\bmy^t\mid \bmu^t)$ and the channel encoder $p_{\btheta}(\bmu^t\mid \bms^t)$. Here one can ask: \emph{how does receiver obtain the true transmit signal $\bmu^t$ to learn channel distribution?} We propose that the receiver also have a transmitter model (including the semantic and channel encoders) that generates the transmit signal $\bmu^t$. Hence, it is essential the receiver's NN model matches accurately with the transmitter model, if the receiver design can correct the errors in the channel and lead to higher semantic reliability.

Assume that, for every semantic state description, there is an optimal action distribution (receiver policy) from which the actions are sampled. This is represented as $\pi_{\theta}^{*}(\bma^t\mid \bmzh^t)$, which represents the implicit semantics that entails from the state description $\bmz^t$. The cross entropy loss for the pair $(\bmz^t,\bmzh^t)$ by using any policy $\pi_{\theta}$ at the receiver can be written as:
\beq
\mL_{CE}(t) = -\sum\limits_{\bma^t} \pi_{\theta}^{*} \log \pi_{\theta} 
\vspace{-2mm}\eeq
Minimizing $\mL_{CE}$ is equivalent to minimizing the KLD between $\pi_{\theta^\ast}$ and $\pi_{\theta}$. Here, at the receiver, we assume that, it is aware of the optimal $\pi_{\theta}^*(\bma^t\mid\bmz^t)$, which is computed using the transmitter model available at its end. The average cross-entropy loss, which can be interpreted as the semantic effectiveness measure can be written as:
\beq
\bar{\mL}_{CE}(t) = -\sum\limits_{\bmz^t,\bmzh^t}\sum\limits_{\bma^t} p(\bmz^t,\bmzh^t)KLD(\pi_{\theta}^{*}\mid\mid \pi_{\theta}),
\eeq
where $p(\bmz^t,\bmzh^t) = p(\bmz^t)p(\bmzh^t\mid \bmz^t) = p(\bmz^t) \int p(\bmzh^t\mid \bmsh^t,[\bmzh^{t-1}])p(\bmsh^t\mid \bmu^t)ds du$.

Now, we have the necessary ingredients to formulate the problem at the receiver side, which can be written as:
\beq
\begin{aligned}
 p(\bmz^t\mid \bmsh^t)^{\ast} = &\argmax\limits_{p(\bmz^t\mid \bmsh^t)} \bar{\mL}_{CE}(t) \\
\mbox{s.t.}\,\, &p\left(E_t\left({\bmz}^t,{\widehat{\bmz}}^t \right) < \delta \right) \geq 1-\epsilon.
\end{aligned}
\eeq
Further, we obtain the augmented Lagrangian as follows:
\beq
\begin{aligned}
 \mP_2:  &\left[p(\bmz^t\mid \bmsh^t)^{\ast}\right] \\  &= \argmax\limits_{p(\bmz^t\mid \bmsh^t)}\bar{\mL}_{CE}(t) + \lambda \left( p\left(E_t\left({\bmz}^t,{\widehat{\bmz}}^t \right) < \delta \right) - (1-\epsilon)\right).
\end{aligned}
\label{eq_P2}
\eeq
The Lagrangian in \eqref{eq_P2} represents the $Q$-function for the receiver side ToM formulation. 
\setlength{\textfloatsep}{0pt}
\begin{algorithm}[t]\scriptsize
\caption{Proposed Solution for Adaptive ToM Collaboration Emergence}\label{alg_ta_CSG}%\vspace{-1mm}
 \textbf{Given:} Randomly initialize $\btheta_i,
\widehat{\btheta}_i, \eta_Q, \eta_{\pi}, \eta_{f} , i \in \{1, \cdots, N\}$.\\  Learning rate $\eta$, Batch size $B$
\begin{algorithmic}[1]
%\vspace{-2mm}\STATE \hspace{0.05cm} 
\vspace{-0mm}\For{each round}
\vspace{-0mm}\For{$i\in \{s,r\}$}
\State Initialize replay buffer $\mD \leftarrow \emptyset$
\While{train agent $i$}
\Repeat
\State Agents sample actions according to equation \eqref{eq_sample_actions}
\State Agents update their beliefs
\State Agents update their estimation of partners’ beliefs
\Until{game ends}
\State Update $\mD$ with new trajectory \\
\State Sample $B$ trajectories $\{(\omega_{1:N} , \bms_{0:T} , \bma_{0:T} , r_{0:T} )$
\State $y_i^{t,(k)} = r_i^{t,(k)} + \gamma \max_{\bms\in \mA_s} Q_{\bthetah_i}(\bms,d^{t+1,(k)},\bomega^{(k)},\widehat{b})$
\State $L^Q =\sum_{t,k} \abs{Q_{\bthetah_i}(\bms,d^{t+1,(k)},\bomega^{(k)},\widehat{b}) - y_i^{t,(k)}}^2$.
\State $L^{\pi} = \sum_{t,k} \mbH(\widehat{\pi}_s(\bma^t\mid d^{t+1,(k)},\bomega^{(k)}),\widehat{\pi}_s(\bma^t\mid)) $
\State $L^f = KL(\widehat{b}_{-i} \mid\mid f_{-i}(\widehat{b}_{-i},\widehat{a}_{-i}))$
\State $\btheta_i \leftarrow \btheta_i - \nabla_{\btheta_i}(\eta^QL^Q+\eta^{\pi}L^{\pi}+\eta^fL^f)$

\vspace{-0mm}\State Periodically update $\bthetah_i \leftarrow \btheta_i$ for Q-learning
\EndWhile
\EndFor
\EndFor
\end{algorithmic}
\label{algo1}  
\vspace{-0mm}\end{algorithm}
\vspace{0mm}

\subsubsection{ToM Solution using centralized training}
The algorithm details are provided in Algorithm~\ref{alg_ta_CSG}. During centralized training, transmitter needs to learn three components, $Q_s, \widehat{\pi}_{r}, f_r$. Similarly, the receiver learns $Q_r, \widehat{\pi}_{s}, f_s$. The ToM in Algorithm.~\ref{alg_ta_CSG} involves three components and are inspired from the NN architecture in \cite{YuanArxiv2021}. $Q_{\bthetah_i}$ is learned using deep Q-learning. The belief network whose lose function is described as $L^{\pi}$ is implemented as a gated recurrent NNs (GRNN) \cite{ChungArxiv2014}. Perception network represented by $\widehat{b}_r(\bomega_{s,i})$ is also implemented as GRNN.

\vspace{-1mm}\subsection{Training}
\vspace{-1mm}
The NN training is jointly performed at the transmitter and receiver using Algorithm \ref{alg_ta_CSG}. The transmitter is assumed to be unaware of the channel information and receiver action distribution. Despite the initial training overhead, the effective throughput remains unaffected as the transmitter NN model dynamically adjusts its parameters based on the two-level feedback information from the receiver. This adaptive approach ensures generalizability and practical feasibility of the SC systems, outperforming state-of-the-art methods \cite{XieTSP2021, FarshbafanArxiv2022} in various wireless channel conditions and tasks.

%At receiver side, it

%1) Perfect knowledge of opposite side NN model. At high SNR, feedback is not useful. Represents and upper bound on the performance for other cases. 
%2) NN Model mismatches. At high SNR, receiver action policy need not be optimal. Feedback leads to corrective action via 

\section{Simulation Results}

In this section, we comprehensively evaluate the proposed ToM-based SC system, assessing its performance based on metrics such as spectral efficiency, semantic reliability, and ability to adapt quickly to changing tasks. We begin with a dataset $\mX = \{\bmx_s\}_{s=1}^S$ of $S$ samples where each sample $\bmx_s$ consists
of $N$ stationary time-series $\bmx_s = \{\bmx_{s,1},\bmx_{s,2},\cdots,\bmx_{s,{N}}\}$ across time-steps $t = \{1, \cdots, T\}$. This dataset represent the states observed at the expert agent. We denote the
$t$-th time-step of the $i$-th time-series of $\bmx_s$ as $\bmx_{s,i}^t \subset \mR^D$. We consider an SCM captured by an associated DAG $\mG_s^{1:T} = \{\mV_s^{1:T},\mE_{s}^{1:T}\}$. underlying the generative process
of each sample. The SCM's endogenous (observed)
variables are vertices $v_{s,i}^{t}\in \mV_s^{1:T}$ for each time-series $i$ and each time-step $t$.  Every set of incoming
edges to an endogenous variable defines inputs to a deterministic function $g^t_{s,i}$ which determines that
variable’s value. Using GNN, our learned NN is assumed to follow the generation of $\bmx_s^t$ as $
\bmx_s^{t+1} = \bmg_s(\bmx_s^{\leq t},\mG_s, \btheta,\bma^t) + \bmv_s^{t+1}.$ We selected a multi-dimensional discrete Gaussian action distribution with its mean value proportional to the number of parents of any entity part of $\bmz^t$. This choice ensures that the vector of actions (with the same dimension as $\bmz^t$) are derived directly from the causal state rather than from the actual observations $\bmo^t$, which may contain variables irrelevant to the receiver actions.

For the constellation, we used BPSK and the single antenna channel between the nodes are chosen as $\mathcal{CN}(0,1)$. The noise variance is adjusted based on the received SNR value. 

In Fig.~\ref{Fig_Theorem_Proving1}, we validate the spectral efficiency of the proposed ToM-based SC system and compare it with 1) SC without ToM (our algorithm without the belief and perception components) and conventional wireless schemes (without any causal state extraction) with repetition coding scheme or HARQ schemes. The figure shows that inducing ToM at the communicating nodes provides better spectral efficiency at low to mid SNRs where the channel errors are predominant. The saturated values at high SNRs are also higher for SC schemes than conventional wireless due to the transmission efficiency arising from capturing semantics, resulting in a reduced number of redundant bit transmissions.

In Figure \ref{Fig_SR_tasks}, we demonstrate the superior performance of our ToM method in terms of semantic reliability compared to several benchmarks. These benchmarks include: 1) causal reasoning-based SC without ToM \cite{ChristoTWCArxiv2022}, 2) imitation learning based implicit reasoning based SC proposed in \cite{XiaoJSAC2023}, and classical AI-based wireless systems that do not extract causal relations \cite{BourtsoulatzeTCCN2019} and utilize deep joint-source channel coding approach. The figure clearly illustrates that our proposed ToM method can dynamically adapt to task changes, resulting in higher semantic reliability than the other benchmark methods.

Figure.~\ref{Fig_SE_samples} illustrates that the proposed ToM-based SC system dynamically adjusts its NN parameters with fewer communication samples, resulting in higher semantic effectiveness. The action distribution changes around $3K$ samples, indicating a task change. However, state-of-the-art SC systems that do not incorporate ToM-based reasoning struggle when the task changes.
\begin{figure}[t]\vspace{-2mm}
%\centerline{\includegraphics[width=8.3cm,height=5cm]{SSP_NMSE_GAMP_IF_SAVE}}
%\centerline{\includegraphics[width=9cm]{NMSE_EM_LAMP_SBL}}
\centerline{\includegraphics[width=3.5in,height=1.8in]{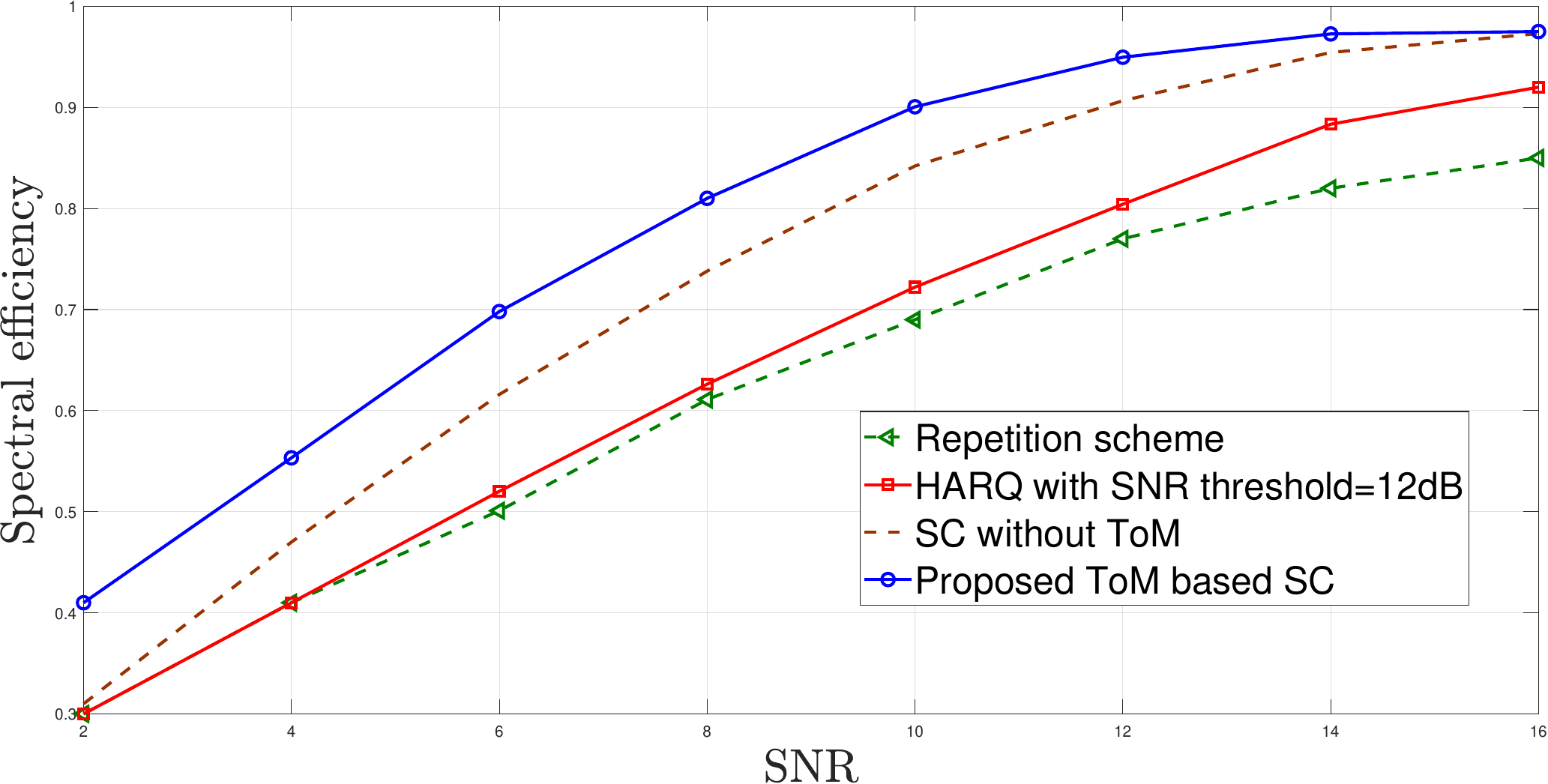}} \vspace{-1mm}
\caption{\scriptsize Spectral efficiency vs SNR for different SC schemes.}
\label{Fig_Theorem_Proving1}
\vspace{-2mm}\end{figure}
\begin{figure}[t]\vspace{-2mm}
%\centerline{\includegraphics[width=8.3cm,height=5cm]{SSP_NMSE_GAMP_IF_SAVE}}
%\centerline{\includegraphics[width=9cm]{NMSE_EM_LAMP_SBL}}
\centerline{\includegraphics[width=3.5in,height=2.1in]{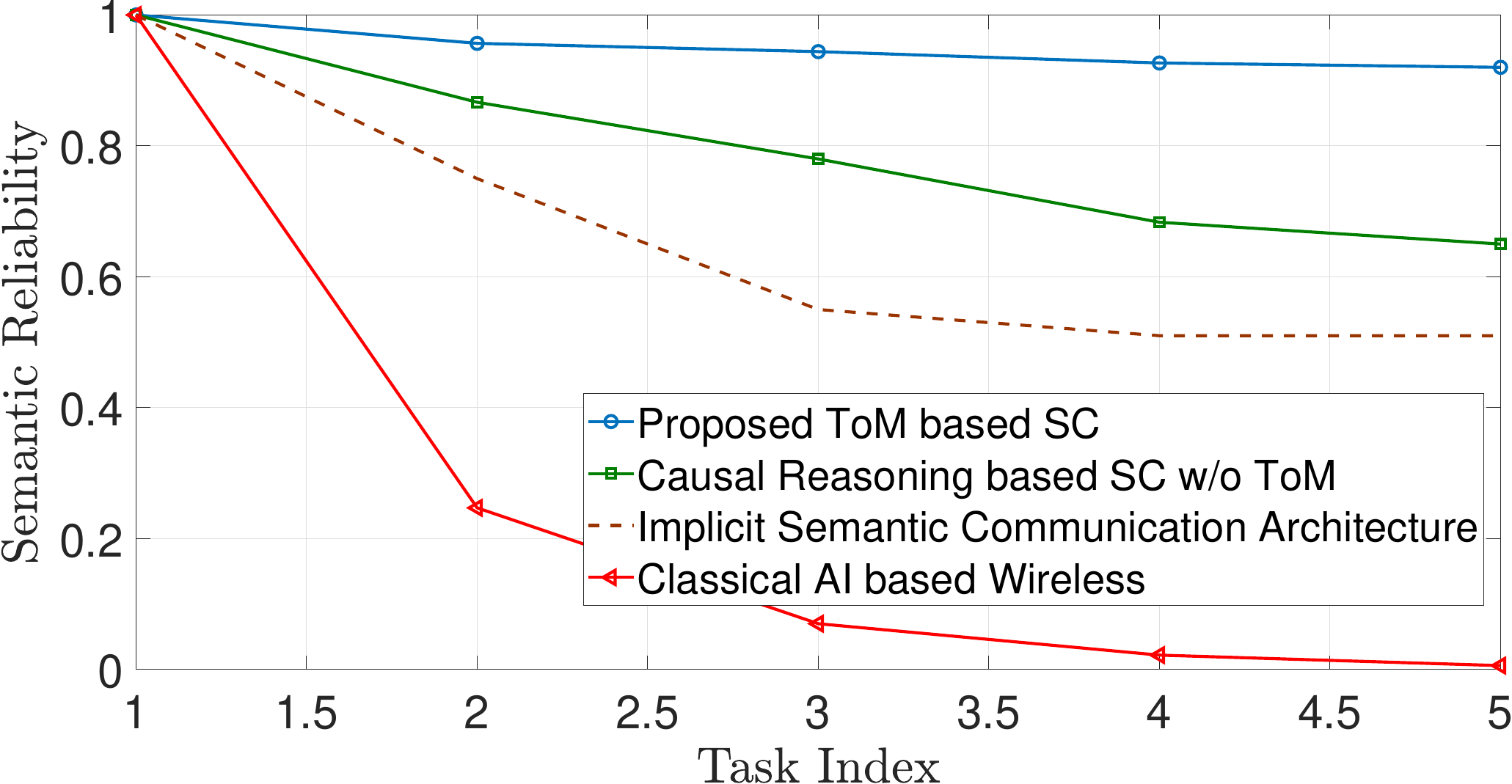}} \vspace{-1mm}
\caption{\scriptsize Semantic reliability vs SNR varying tasks.}
\label{Fig_SR_tasks}
\vspace{-0mm}\end{figure}
% Why transmitter not send directly the actions (optimal) ? When channel is bad, quality of reception is affected? 
\begin{figure}[h]\vspace{-3mm}
%\centerline{\includegraphics[width=8.3cm,height=5cm]{SSP_NMSE_GAMP_IF_SAVE}}
%\centerline{\includegraphics[width=9cm]{NMSE_EM_LAMP_SBL}}
\centerline{\includegraphics[width=3.5in,height=1.9in]{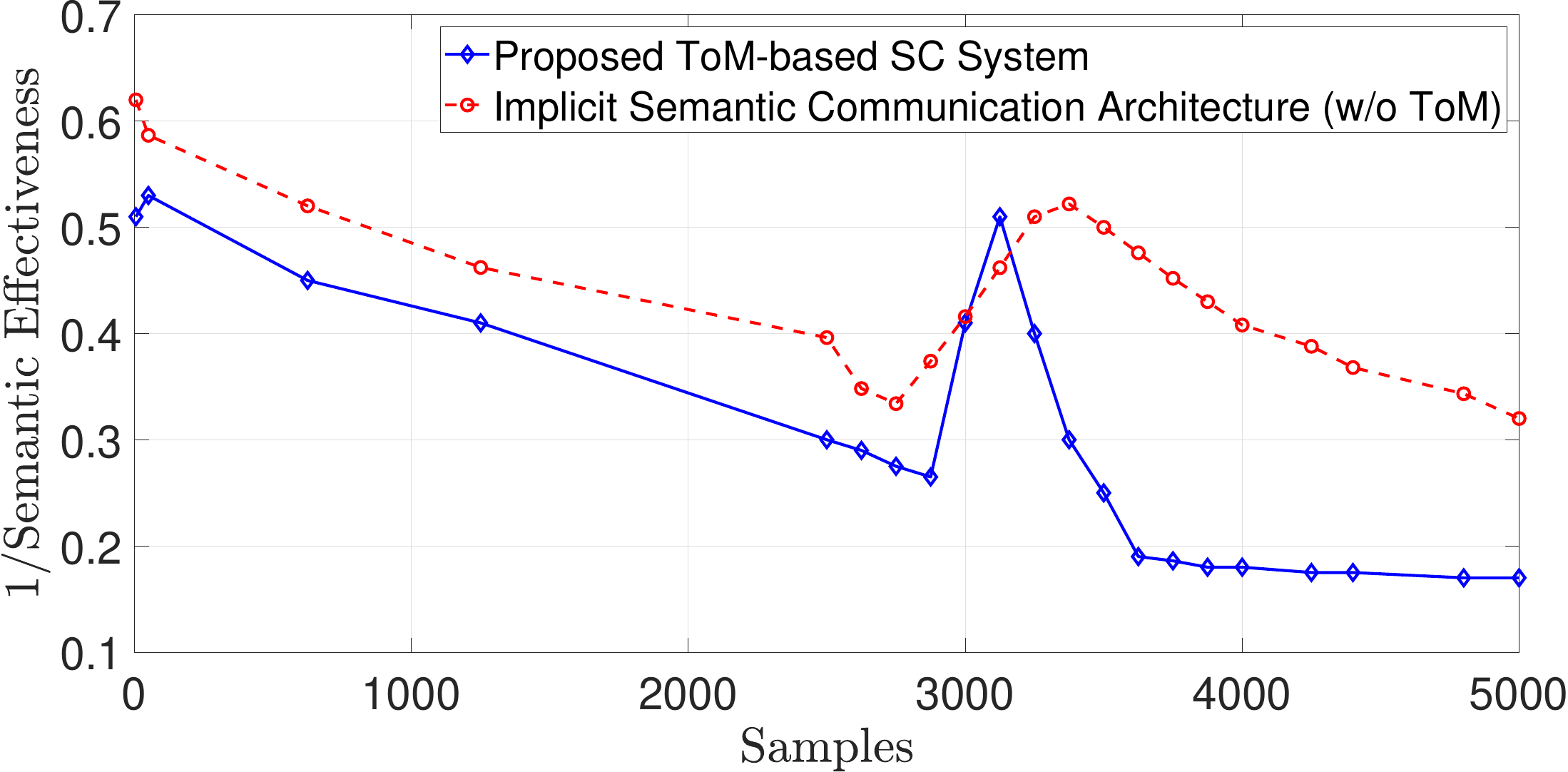}} \vspace{-1mm}
\caption{\scriptsize Semantic effectiveness vs number of samples, with task changing around $3K$ samples. }
\label{Fig_SE_samples}
\vspace{-2mm}\end{figure}

\vspace{-2mm}\section{Conclusion}
 \vspace{-0mm}

In this paper, we have presented a new vision of semantic communication systems that relies upon ToM as a machinery to enable dynamic updates of the NN components at the transmit and receive side. We have formulated a ToM based joint transmit and receive design to learn an optimal representation space for transmission and extracting the semantic state description at the listener. The optimized AI components at the transmit side herein can be dynamically fine-tuned using the semantic level feedback from the receive side. %We have also proposed a reasoning component that lets the compositional language extract complex event descriptions at the listener using the chain of thought type of reasoning. 
The causal discovery component at the transmitter is implemented using the GNN and is learned as a directed acyclic graph. Simulation results have demonstrated superiority of our proposed SC in terms of improving the communication efficiency (minimal transmission) and reliability compared to classical communication and state of the art semantic communication. 

\vspace{-2mm}\bibliographystyle{IEEEbib}
\bibliography{refs,semantics_ref}

%\bibliographystyle{IEEEbib}
%\bibliography{semantic_refs}
%\vspace{-1mm}
\end{document}

%% file: defines.tex
\newcommand{\beq}{\begin{equation}}
\newcommand{\eeq}{\end{equation}}

\newcommand{\mZ}{\mathcal{Z}}

\DeclareMathOperator*{\argmax}{arg\,max}
\DeclareMathOperator*{\argmin}{arg\,min}
\def\adots{\mathinner{\mskip0mu\raise0pt\vbox{\kern7pt\hbox{.}}\mskip3mu
          \raise4pt\hbox{.}\mskip3mu\raise8pt\hbox{.}\mskip0mu}}

\usepackage{bm}

\newcommand{\bmx}{{\bm x}}
\newcommand{\bmy}{{\bm y}}
\newcommand{\bmv}{{\bm v}}

\newcommand{\bmxh}{\widehat{\bmx}}
\newcommand{\mS}{\mathcal{S}}
\newcommand{\mX}{\mathcal{X}}

\newcommand{\mD}{\mathcal{D}}
\newcommand{\mbS}{\mathbb{S}}
\newcommand{\mbH}{\mathbb{H}}
\newcommand{\mbE}{\mathbb{E}}
\newcommand{\mP}{\mathcal{P}}
\newcommand{\mO}{\mathcal{O}}
\newcommand{\mA}{\mathcal{A}}
\newcommand{\mR}{\mathcal{R}}
\newcommand{\mE}{\mathcal{E}}
\newcommand{\mV}{\mathcal{V}}
\newcommand{\mN}{\mathcal{N}}

\newcommand{\mM}{\mathcal{M}}

\newcommand{\bmzh}{\widehat{\bmz}}

\newcommand{\bmo}{{\boldsymbol {o}}}

\newcommand{\bmn}{{\bm n}}
\newcommand{\bma}{{\bm a}}

\newcommand{\bomega}{\boldsymbol{\omega}}
\newcommand{\btheta}{\boldsymbol{\theta}}
\newcommand{\bthetah}{\widehat{\btheta}}

\newcommand{\mbI}{\mathbb{I}}

%% file: main.bbl
\begin{thebibliography}{10}

\bibitem{ChaccourITJ2022}
C.~Chaccour, M.~N. Soorki, W.~Saad, M.~Bennis, and P.~Popovski,
\newblock ``{Can Terahertz Provide High-Rate Reliable Low-Latency Communications for Wireless VR},''
\newblock {\em IEEE Internet of Things Journal}, vol. 9, no. 12, pp. 9712--9729, Jun. 2022.

\bibitem{ChaccourArxiv2022}
C.~Chaccour, W.~Saad, M.~Debbah, Z.~Han, and H.~V. Poor,
\newblock ``{Less Data, More Knowledge: Building Next Generation Semantic Communication Networks},''
\newblock {\em arXiv preprint arXiv:2211.14343}, Nov. 2022.

\bibitem{StrinatiComNetworks2021}
E.~C. Strinati and S.~Barbarossa,
\newblock ``{6G Networks: Beyond Shannon Towards Semantic and Goal-Oriented Communications},''
\newblock {\em Computer Networks, Elsevier}, vol. 190, May 2021.

\bibitem{UysalIN2022}
E.~Uysal, O.~Kaya, A.Ephremides, J.~Gross, M.~Codreanu, P.~Popovski, M.~Assad, G.~Liva, A.~Munari, B.~Soret, and T.~Soleymani,
\newblock ``{Semantic Communications in Networked Systems: A Data Significance Perspective},''
\newblock {\em IEEE Network}, vol. 36, no. 4, pp. 233--240, Oct. 2022.

\bibitem{XieTSP2021}
H.~Xie, Z.~Qin, G.~Y. Li, and B-H. Juang,
\newblock ``{Deep Learning Enabled Semantic Communication Systems},''
\newblock {\em IEEE Transactions on Signal Processing}, vol. 69, pp. 2663--2675, Apr. 2021.

\bibitem{FarshbafanArxiv2022}
M.~K. Farshbafan, W.~Saad, and M.~Debbah,
\newblock ``{Curriculum Learning for Goal-Oriented Semantic Communications with a Common Language},''
\newblock {\em arXiv preprint arXiv:2111.08051}, Feb. 2022.

\bibitem{Sana2022}
M.~Sana and E.~C. Strinati,
\newblock ``{Learning Semantics: An Opportunity for Effective 6G Communications},''
\newblock in {\em Proceedings of IEEE 19th Annual Consumer Communications and Networking Conference (CCNC)}, 2022, pp. 631--636.

\bibitem{CalvanesePatent2020}
E.~Calvanese~Strinati,
\newblock ``{Semantic Data Exchange System with A Semantic Retransmission Channel },''
\newblock {\em French Patent FR3111199, European Patent EP3923138, US patent US11706291}, 24th, Aug. 2020.

\bibitem{Petar2020}
P.~Popovski, O.~Simeone, F.~Boccardi, D.~G{\"u}nd{\"u}z, and O.~Sahin,
\newblock ``{Semantic-Effectiveness Filtering and Control for Post-5G Wireless Connectivity },''
\newblock {\em Journal of the Indian Institute of Science}, vol. 100, no. 2, pp. 435--443, Apr. 2020.

\bibitem{ChristoTWCArxiv2022}
C.~K. Thomas and W.~Saad,
\newblock ``{Neuro-Symbolic Causal Reasoning Meets Signaling Game for Emergent Semantic Communications},''
\newblock {\em IEEE Transactions on Wireless Communications}, Oct. 2023.

\bibitem{YuanArxiv2021}
L.~Yuan, Z.~Fu, L.~Zhou, K.~Yang, and S-C Zhu,
\newblock ``{Emergence of Theory of Mind Collaboration in Multiagent Systems},''
\newblock {\em arXiv preprint arXiv:2110.00121}, pp. 424--438, 2021.

\bibitem{ScarselliGNN2008}
F.~Scarselli, M.~Gori, A.~C. Tsoi, M.~Hagenbuchner, and G.~Monfardini,
\newblock ``{The Graph Neural Network Model},''
\newblock {\em IEEE Transactions on Neural Networks}, vol. 20, no. 1, pp. 561--80, Dec. 2008.

\bibitem{DudekAAAI2009}
G.~Dudek and N.~Roy,
\newblock ``{Multi-Robot Rendezvous in Unknown Environments},''
\newblock in {\em Proceedings of the AAAI National Conference Workshop on Online Search}, Providence, Rhode Island, USA, 2009, pp. 22--29.

\bibitem{BareinboimACM2020}
E.~Bareinboim, J.~D. Correa, D.~Ibeling, and T.~Icard,
\newblock ``{On Pearl’s Hierarchy and the Foundations of Causal Inference},''
\newblock in {\em Probabilistic and Causal Inference: The Works of Judea Pearl (ACM Books)}, 2020.

\bibitem{HafnerPMLR2019}
D.~Hafner, T.~Lillicrap, T.~Fischer, T.~Villegas~D. Ha, H.~Lee, and J.~Davidson,
\newblock ``{Learning Latent Dynamics for Planning from Pixels},''
\newblock {\em Proceedings of the 36th International Conference on Machine Learning (PMLR)}, vol. 97, Jun. 2019.

\bibitem{ChristoJSAITArxiv2023}
C.~K. Thomas, W.~Saad, and Y.~Xiao,
\newblock ``{Causal Semantic Communication for Digital Twins: A Generalizable Imitation Learning Approach},''
\newblock {\em arXiv e-prints. Apr:arXiv-2304}, 2023.

\bibitem{GrangerJES1969}
C.~W.~J. Granger,
\newblock ``{Investigating Causal relations by Econometric Models and Cross-spectral Methods},''
\newblock {\em Journal of the Econometric Society}, pp. 424--438, 1969.

\bibitem{FoxAIR2012}
C.~W. Fox and S.~J. Roberts,
\newblock ``{A Tutorial on Variational Bayesian Inference},''
\newblock {\em Artificial intelligence review}, vol. 38, pp. 85--95, 2012.

\bibitem{WarglienSynthese2013}
M.~Warglien and P.~G{\" a}rdenfors,
\newblock ``{Semantics, Conceptual Spaces, and the Meeting of Minds.},''
\newblock {\em Synthese}, vol. 190, no. 12, pp. 2165--2193, 2013.

\bibitem{YuanNeurIPS2019}
L.~Yuan, Z.~Fu, L.~Zhou, K.~Yang, and S-C Zhu,
\newblock ``{Emergence of Theory of Mind Collaboration in Multiagent Systems},''
\newblock in {\em 33rd Conference on Neural Information Processing Systems (NeurIPS)}, Vancouver, Canada, Dec. 2019.

\bibitem{TungJSAIT2022}
T.~Y. Tung, D.~B. Kurka, M.~Jankowski, and D.~G{\"u}nd{\"u}z,
\newblock ``{DeepJSCC-Q: Constellation Constrained Deep Joint Source-Channel Coding},''
\newblock {\em IEEE Journal on Selected Areas in Information Theory}, vol. 3, pp. 720 -- 731, 2022.

\bibitem{ChungArxiv2014}
J.~Chung, C.~Gulcehre, K.~Cho, and Y.~Bengio,
\newblock ``{Empirical Evaluation of Gated Recurrent Neural Networks on Sequence Modeling},''
\newblock {\em arXiv preprint arXiv:1412.3555}, 2014.

\bibitem{XiaoJSAC2023}
Y.~Xiao, Z.~Sun, G.~Shi, and D.~Niyato,
\newblock ``{Imitation Learning-based Implicit Semantic-aware Communication Networks: Multi-layer Representation and Collaborative Reasoning},''
\newblock {\em IEEE Journal on Selected Areas in Communications}, vol. 43, no. 1, Mar. 2023.

\bibitem{BourtsoulatzeTCCN2019}
E.~Bourtsoulatze, D.~B. Kurka, and D.~G{\"u}nd{\"u}z,
\newblock ``{Deep Joint Source-Channel Coding for Wireless Image Transmission},''
\newblock {\em IEEE Transactions on Cognitive Communications and Networking}, vol. 5, no. 3, pp. 567--579, 2019.

\end{thebibliography}
